\documentclass[preprint,aps,prc,showpacs]{revtex4}

\usepackage{graphicx}% Include figure files
\usepackage{bm}% bold math

\begin{document}

\title{Antiproton-nucleus collisions simulation within a kinetic
       approach with relativistic mean fields.}

\author{A.B. Larionov$^{1,2}$\footnote{Corresponding author.\\ 
        E-mail address: larionov@fias.uni-frankfurt.de}, 
        I.A. Pshenichnov$^{1,3}$, I.N. Mishustin$^{1,2}$, 
        and W. Greiner$^1$}

\affiliation{$^1$Frankfurt Institute for Advanced Studies, J.W. Goethe-Universit\"at,
             D-60438 Frankfurt am Main, Germany\\ 
             $^2$Russian Research Center Kurchatov Institute, 
             123182 Moscow, Russia\\
             $^3$Institute for Nuclear Research, Russian Academy of Science, 
             117312 Moscow, Russia}

\date{\today}

\begin{abstract}
The Giessen Boltzmann-Uehling-Uhlenbeck transport model with relativistic
mean fields is used to simulate $\bar p$-nucleus collisions. 
Antiproton absorption cross sections and 
momentum distributions of annihilation products are calculated by varying 
the $\bar p$ coupling strength to the mean meson fields.
Parameters of the antiproton-nucleus optical potential are extracted from
the comparison of the model calculations with experimental data.  
\end{abstract}


\maketitle

The real and imaginary parts of the antiproton optical potential are key quantities
which determine $\bar p$-nucleus scattering at low 
\cite{Garreta1984266,Garreta:1984rs,Nakamura:1984xw} 
and intermediate \cite{Abrams:1972ab} energies, as well as the existence of 
deeply bound $\bar N$-nucleus states \cite{Wong:1984fy,Baltz:1985zz,Buervenich:2002ns,%
Mishustin:2004xa,Friedman:2005ad,Friedman:2007zz}. The $\bar p N$ interaction
in nuclear matter is, moreover, a challenging problem by itself, since it is subject to 
strong in-medium modifications. Indeed, in the simplest $t\rho$-approximation the real part
of the $\bar p$ optical potential is repulsive and about 100 MeV high 
for the $\bar N N$ amplitude $t$ taken at threshold \cite{Batty:1997zp}, 
while $\bar p$-atomic phenomenology requires a strong attraction 
\cite{Friedman:2005ad,Friedman:2007zz,Batty:1997zp}.

Usually the nuclear part of the 
$\bar p$ optical potential is parameterized in a Woods-Saxon (WS) form
\begin{equation}
   V_{\rm opt} \simeq -\frac{V_0}{\exp\left(\frac{r-R_R}{a_R}\right)+1}
      -\frac{i\,W_0}{\exp\left(\frac{r-R_I}{a_I}\right)+1}~.   \label{Vopt_WS}
\end{equation}
In spite of several previous attempts to fix the $\bar p$ optical potential,     
considerable ambiguity still remains in its parameters. 
The angular distributions of elastically scattered antiprotons favour a shallow 
real part $V_0=0\div70$ MeV and a deep imaginary part $W_0=70\div150$ MeV in the interior 
of a nucleus \cite{Nakamura:1984xw,Garreta1984266,Garreta:1984rs,Batty:1984kw,%
Kronenfeld1984525,Batty:1984ra,Dalkarov:1986tu,Friedman:1986nf}. 
The Glauber and optical model calculations for the $\bar p$ absorption on
nuclei \cite{Kondratyuk:1986cq,Lenske:2005nt} also assume a negligibly small real part 
of $V_{\rm opt}$ and a strongly absoptive imaginary part.   
At the same time, the most recent combined analysis \cite{Friedman:2005ad} 
of the X-ray transitions in antiprotonic atoms and of the radiochemical data has 
produced a deep real part $V_0=110$ MeV and an imaginary part $W_0=160$ MeV.

The $\bar p$ optical potentials from elastic scattering and $\bar p$-atomic data
are well determined only at the extreme periphery of a nucleus, corresponding to less 
than 10\% of the central density \cite{Garreta:1984rs,Friedman:2005ad}. 
On the other hand, $\bar p$ production in proton-nucleus and nucleus-nucleus 
collisions probes the $\bar p$ potential deeply inside the nucleus and favours 
$V_0=100\div200$ MeV consistent with a dispersion relation 
between real and imaginary parts of $V_{\rm opt}$ \cite{Teis:1994ie}.
However, the microscopic transport analysis of Ref. \cite{Teis:1994ie} is also sensitive
to rather uncertain in-medium elementary $\bar p$-production cross sections close to threshold.

The purpose of this work is to extract the information on a $\bar p$
optical potential from the data on $\bar p$ absorption cross section
on nuclei \cite{Nakamura:1984xw,Abrams:1972ab,Denisov:1973zv,Carroll:1978hc} and from the data
on inclusive pion and proton production from low-energy $\bar p$ annihilation
in nuclei \cite{Mcgaughey:1986kz}. Here, the absorption means the removal of a $\bar p$
from a beam caused by the annihilation, (in)elastic scattering and charge exchange
reactions on individual nucleons. The diffractive elastic scattering on a nucleus
as a whole is excluded from the absorption cross section. Since the absorption requires
at least one $\bar p N$ collision, it is sensitive to the $\bar p$ optical potential
in a deeper region of a nucleus with respect to the case of diffractive elastic scattering.
Indeed, the annihilation of 180 MeV antiprotons takes place at about half-density radius
\cite{Clover:1982qq,Cugnon:1986tx}. By the same argument, the proton and pion production from 
$\bar p$ annihilation on nuclei should also probe the $\bar p$ optical potential at about 
half-density radius. 

In our calculations, we have applied the Giessen Boltzmann-Uehling-Uhlenbeck (GiBUU)
model \cite{GiBUU}. 
This model solves a system of semiclassical kinetic equations for baryons,
antibaryons and mesons coupled via collision terms and mean fields. 
The phase space distribution function of every particle species is projected on a set 
of point-like test particles.
The coordinates ${\bf r}_j$ and space components of the kinetic four-momentum 
$p_j^{\star \mu}$ of a baryon ($j=B$) or an antibaryon ($j=\bar B$) test particle are 
propagated in time according to the following Hamiltonian-like equations 
(c.f. \cite{Larionov:2007hy,Larionov:2008wy,Gaitanos:2007mm} and refs. therein):
\begin{eqnarray}
& & \dot{\bf r}_j =  \frac{{\bf p}_j^\star}{p_j^{\star 0}}~,                     \label{rDot} \\
& & \dot{p}^{\star k}_j = \frac{p_{j\mu}^\star}{p_j^{\star 0}} F_j^{k\mu}
                              + \frac{m_j^\star}{p_j^{\star 0}} 
                                \frac{\partial m_j^\star}{\partial r_k}     \label{pStarDot}
\end{eqnarray}
with $k=1,2,3$ and $\mu=0,1,2,3$. The particles are assumed to be on the effective mass 
shell, $p_j^{\star 0}=\sqrt{({\bf p}_j^\star)^2 + (m_j^\star)^2}$. The kinetic four-momentum 
is defined as $p_j^{\star \mu} \equiv p_j^\mu - V_j^\mu$, where $p_j^\mu$ is a canonical four-momentum
and 
\begin{equation}
   V_j^\mu = g_{\omega j} \omega^\mu + g_{\rho j} \tau^3 \rho^{3\mu} 
         + \frac{e}{2} (B_j+\tau^3) A^\mu                                \label{V^mu}
\end{equation}
is a vector field with $\tau^3=+1$ for $p$ and $\bar n$, $\tau^3=-1$ for $\bar p$ and $n$, and
$B_B=+1$, $B_{\bar B}=-1$ being the baryon number. 
The field tensor in the r.h.s. of (\ref{pStarDot}) is defined as
$F_j^{\nu\mu} \equiv \partial^\nu V_j^\mu - \partial^\mu V_j^\nu$.
The effective mass $m_j^\star$ is expressed in terms of a scalar potential $S_j=g_{\sigma j} \sigma$
as $m_j^\star= m_j + S_j$. 
Mesonic mean fields included into the model are $(I,S)=(0,1)$ $\omega$,
(1,1) {\boldmath ${\mathbf \rho}$  \unboldmath}, and (0,0) $\sigma$ with $g_{\omega j}$, $g_{\rho j}$, 
and $g_{\sigma j}$ being the respective coupling constants. The time component of the electromagnetic 
field, i.e. Coulomb potential, is also taken into account.
The mesonic mean fields and Coulomb potential are calculated by solving the field
equations with source terms provided by the currents and scalar density of test
particles on the basis of a Relativistic Mean Field (RMF) model 
\cite{Larionov:2007hy,Larionov:2008wy,Gaitanos:2007mm}. Initial positions and momenta
of the test particles in the ground state nuclei are chosen randomly according to 
the spatial density distributions and the local Fermi momentum distribution. 
The proton and neutron densities are taken in a WS form consistent with 
a Skyrme Hartree-Fock systematics \cite{Lenske_private}.

The RMF and optical models can be related via a Schr\"odinger equivalent potential 
\cite{Friedman:2005ad,Bouyssy:1982cn}:
\begin{equation}
   \mbox{Re}(V_{\rm opt}) = S_j + V_j^0 + \frac{S_j^2-(V_j^0)^2}{2m_j} 
                          + \frac{V_j^0}{m_j}E_{\rm lab},                   \label{Re_Vopt}
\end{equation}
where $E_{\rm lab}=\sqrt{p_{\rm lab}^2+m_j^2}-m_j$ is the kinetic energy of a beam particle 
far away from a nucleus. The real part of the $\bar p$ optical potential becomes deeper 
with increasing $E_{\rm lab}$ due to the negative vector potential, in distinction to the 
real part of the proton optical potential \cite{Bouyssy:1982cn}.

The $\sigma$-, $\omega$- and $\rho$-nucleon coupling constants have been taken 
from the NL3 model \cite{Lalazissis:1996rd} providing a very good description of the ground
states for both spherical and deformed nuclei. The meson-antinucleon coupling constants
are quite uncertain. Following \cite{Mishustin:2004xa}, we introduce
$g_{\omega \bar N} = -\xi g_{\omega N},~g_{\rho \bar N} = \xi g_{\rho N},~
g_{\sigma \bar N} = \xi g_{\sigma N}$, where $0 < \xi \leq 1$ is an adjutsable parameter.
The case of $\xi=1$ corresponds to the $G$-parity transformed nucleon fields. 
For $\xi=1$, neglecting the Coulomb field, the value of the $\bar p$ vector potential 
in nuclear matter at the saturation density $\rho_0=0.148$ fm$^{-3}$ is $V_{\bar p}^0 = -308$ MeV, 
while the scalar potential is $S_{\bar p} = -380$ MeV. 
This gives an extremely deep real part, $\mbox{Re}(V_{\rm opt}) = -661$ MeV.
Below, we will try to find out the values of $\xi$ which are best suited to describe
$\bar p$ absorption and annihilation data on nuclei.

The antinucleon-nucleon collision terms in kinetic equations describe the elastic scattering,
inelastic production and annihilation processes:
$\bar N N \to \bar N N$ (including charge exchange), $\bar N N \to \bar B B + {\rm mesons}$,
$\bar N N \to {\rm mesons}$. 
The cross sections of these processes are based on the experimental data 
parameterizations \cite{Cugnon:1989,Montanet:1994xu}.
The $\bar N N$ annihilation has been described on the basis of a statistical model 
\cite{Golubeva:1992tr,PshenichnovPhD}.
The annihilation final state includes up to six particles, which are various combinations 
of $\pi$, $\eta$, $\omega$ and $\rho$ mesons.
The annihilation model was originally used in Ref. \cite{Golubeva:1992tr}
for slow antiprotons, but after the proper parameter adjustment the model also
describes successfully pion spectra and multiplicity distributions in $\bar{p}p$ 
annihilation in flight, up to $p_{\rm lab} \simeq 10$ GeV/c \cite{PshenichnovPhD}. 
A more detailed description of the $\bar N N$  collision channels implementation in the GiBUU 
model will be given elsewhere.   

To calculate collision terms, we have used a full ensemble technique of the
test particle method (c.f. \cite{Buss:2006yk} and refs. therein).
The full ensemble technique, in distinction to the cascade-like parallel ensemble 
one, solves  the Boltzmann equation more precisely as the binary collisions 
are better localized. In the $\bar p$ absorption calculation on nuclei, 
we have turned off the two-body collisions of the secondary particles in order to enforce 
a Glauber-type description. 
\begin{figure}
   \begin{center}
      \begin{tabular}{cc}
         \includegraphics[scale = 0.45]{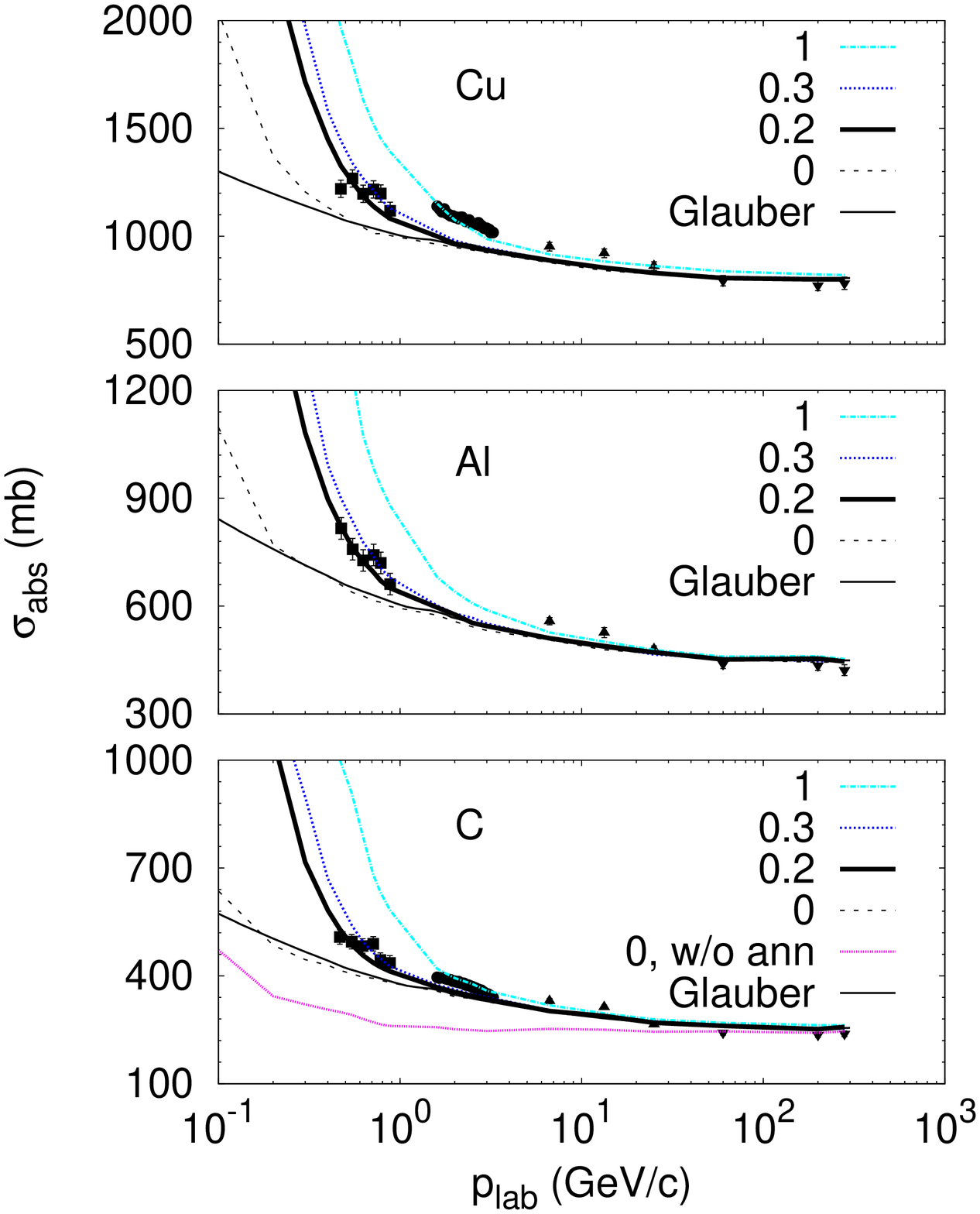} &
         \includegraphics[scale = 0.45]{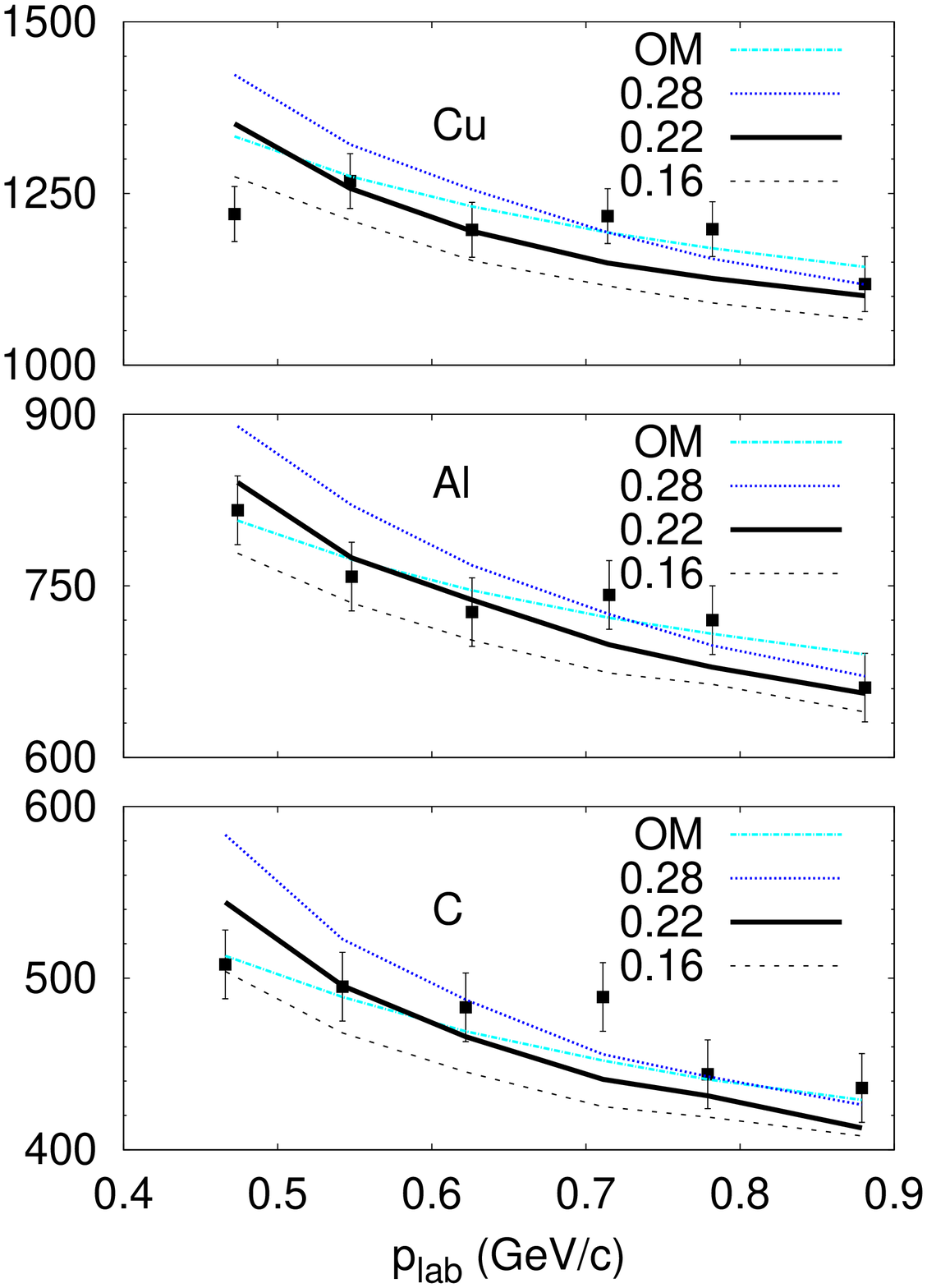} \\
      \end{tabular}

\vspace*{1.5cm}

\caption{\label{fig:sigabs}(Color online) $\bar p$ absorption cross section on 
various nuclei vs the beam momentum. 
The lines marked with the value of a scaling factor $\xi$ show the GiBUU results.
On the left three panels, thin solid lines 
represent the Glauber model calculation, Eq.(\ref{sigabs^Glauber}). 
For the $\bar p$+$^{12}$C system, 
a calculation with $\xi=0$ without annihilation is additionally shown by the dotted line
on the lower left panel.  
Data are from Ref. \cite{Nakamura:1984xw} (filled boxes), Ref. \cite{Abrams:1972ab} 
(filled circles), Ref. \cite{Denisov:1973zv} (filled triangles) and Ref. \cite{Carroll:1978hc}
(filled upside-down triangles).
The right three panels show $\sigma_{\rm abs}$ at lower beam momenta, where the actual 
fit has been done. The optical model (OM) results are from Ref. \cite{Nakamura:1984xw}.}
   \end{center}
\end{figure}

Fig.~\ref{fig:sigabs} shows the $\bar p$ absorption cross section on $^{12}$C,
$^{27}$Al and $^{64}$Cu as a function of the beam momentum. The 
absorption cross section has been computed as
\begin{equation}
   \sigma_{\rm abs}=2\pi\int\limits_0^{b_{\rm max}}\,db\, b\, 
                        P_{\rm abs}(b)~,              \label{sigabs}
\end{equation}
where $b$ is an impact parameter and $P_{\rm abs}(b)$ is the probability of
the $\bar p$ absorption, i.e. suffering at least one (in)elastic scattering 
or annihilation on a nucleon.  The maximum value of the impact parameter, 
$b_{\rm max} \simeq (1.1A^{1/3}+5)$ fm, has been chosen large enough to 
have $P_{\rm abs}(b_{\rm max})=0$ in a GiBUU calculation. Without any mean
field effects, Eq.(\ref{sigabs}) can be reduced to a simple Glauber formula
\cite{Abrams:1972ab,Glauber:1970jm}
\begin{equation}
   \sigma_{\rm abs}^{\rm Glauber}=2\pi\int\limits_0^\infty\,db\, b\,
         [ 1 - \exp( -\overline\sigma_{\rm tot} S(b)) ]~,     \label{sigabs^Glauber}
\end{equation}
where $\overline\sigma_{\rm tot}$ is the isospin-averaged total $\bar p N$ cross section and  
\begin{equation}
   S(b)=2\int\limits_0^\infty\,\rho(\sqrt{b^2+s^2}) ds     \label{S}
\end{equation}
is the nuclear density integral along the straight line trajectory of a projectile
with $\rho(r)$ being the nuclear density at the radial distance $r$. 
The Glauber formula (\ref{sigabs^Glauber}) describes quite well the experimental 
absorption cross section at very high beam momenta. However, there are significant 
deviations of Eq. (\ref{sigabs^Glauber}) from the data at 
$p_{\rm lab} \leq 20$ GeV/c.

At $p_{\rm lab} > 0.4$ GeV/c, the GiBUU calculations with $\xi=0$ are very close to
the Glauber model, as expected. At smaller beam momenta, the attractive Coulomb potential 
increases the absorption cross section with respect to Eq. (\ref{sigabs^Glauber}). 
As one can see from Fig.~\ref{fig:sigabs}, the agreement 
with the data can only be achieved when the mean meson 
fields are introduced. This can be understood as follows: In the calculation without any 
mean field, the beam particles with impact parameters larger than the nuclear radius do not 
experience binary collisions since they propagate along straight-line trajectories.
Turning on the attractive mean field bends trajectories of the beam particles toward the nucleus. 
Thus, the attractive mean field makes a larger
part of the beam flux to experience two-body collisions.

\begin{table}[htb]
\caption{\label{tab:WS} The scaling factor $\xi$ of the antibaryon coupling constants
and $\bar p$ optical potential parameters $V_0$ (MeV), $R_R$ (fm), $a_R$ (fm),
$W_0$ (MeV), $R_I$ (fm), $a_I$ (fm) (see Eq. (\ref{Vopt_WS})),
obtained by fitting the data of Ref. \cite{Nakamura:1984xw} for different nuclei.
The $\chi^2$ values per degree of freedom ($F=5$) and the standard errors of the
scaling factor $\xi$ are given. The errors of the real depth $V_0$ are caused by 
a variation of $\xi$ by one standard error. The errors of all other parameters
are less than 2\% and are not shown.}

\begin{ruledtabular}
\begin{tabular}{ccccccccc}
Nucleus & $\xi$ & $\chi^2/F$ & $V_0$ & $R_R A^{-1/3}$ & $a_R$ & $W_0$ & $R_I A^{-1/3}$ & $a_I$ \\  
\hline
$^{12}$C &0.22$\pm$0.03&2.2&153$\pm$21&1.00&0.63&110&0.97&0.52 \\
$^{27}$Al&0.21$\pm$0.04&1.1&162$\pm$37&1.04&0.64&108&0.99&0.66 \\
$^{64}$Cu&0.21$\pm$0.04&3.3&153$\pm$29&1.09&0.64&103&1.06&0.65 \\
\end{tabular}
\end{ruledtabular}
\end{table}

The sensitivity of the absorption cross section to the $\bar p$ mean field
grows with decreasing beam momentum. Thus we have selected the KEK data 
\cite{Nakamura:1984xw} at $p_{\rm lab}=470\div880$ MeV/c to find the optimum 
value of the parameter $\xi$ for $^{12}$C, $^{27}$Al and $^{64}$Cu targets.
As one can see from Fig.~\ref{fig:sigabs}, $\xi=0.2\div0.3$ provides the best
overall agreement with the data. A stronger attraction, i.e. larger $\xi$,
leads to an overestimation of the absorption cross section at $p_{\rm lab}<1$ GeV/c.
Minimizing $\chi^2$ deviation from six data points for each nucleus results in 
$\xi$ values listed in Table~\ref{tab:WS}. 
The global fit to eighteen data points for all three nuclei produces the scaling
factor $\xi=0.21\pm0.03$ with $\chi^2/F=2.0$ ($F=17$).

The quality of our calculations is visualised in Fig.~\ref{fig:sigabs} 
(right panels), where we also show for a comparison the absorption 
cross sections from the optical model calculations of Ref. 
\cite{Nakamura:1984xw}.
Our absorption cross section drops with increasing beam momentum
somewhat faster than the data do.
The optical model describes the data slightly better. However, 
the optical potential of Ref. \cite{Nakamura:1984xw} has the two
free parameters, $V_0$ and $W_0$, vs only one, i.e. $\xi$ (or, equivalently,
$V_0$) in our model. 
Moreover, the fixed geometrical parameters of the optical potential of 
Ref. \cite{Nakamura:1984xw} are rather arbitrary and, therefore,
can be considered as free parameters too.

We have also extracted the parameters of a $\bar p$ optical potential 
for the best fit values of $\xi$. The real part of $V_{\rm opt}$ has been determined 
from Eq. (\ref{Re_Vopt}) dropping the Coulomb field contribution in $V_{\bar p}^0$. 
The imaginary part has been calculated as
\begin{equation}
   \mbox{Im}(V_{\rm opt}) = -\frac{1}{2} 
                             <v_{\rm rel} \sigma_{\rm tot}^{\rm med}> \rho~, 
                                                    \label{Im_Vopt}
\end{equation}
where the averaging is done with respect to the Fermi momenta of nucleons,
$v_{\rm rel}$ is the relative velocity of an incoming $\bar p$ and a nucleon,
$\sigma_{\rm tot}^{\rm med}$ is the total in-medium $\bar pN$ cross section 
computed taking into account the Pauli blocking of a final nucleon state for the 
(in)elastic scattering contribution, and $\rho$ is the density of nucleons. 
The radial dependencies of the real and imaginary parts, Eqs. (\ref{Re_Vopt}) 
and(\ref{Im_Vopt}), have been approximated by Eq. (\ref{Vopt_WS}) for $E_{\rm lab}=0$.
Resulting WS parameters are listed in Table~\ref{tab:WS}.
The mass dependence of the WS radii $R_R$, $R_I$ appreciably deviates
from a standard $\propto A^{1/3}$ behaviour, which is mostly caused by the
underlying realistic neutron and proton density distributions from a Skyrme 
Hartree-Fock systematics \cite{Lenske_private}. The extraction of the optical 
potential parameters is far from unique.  In particular, the WS depths
are sensitive to the assumed size parameters, and there is no guarantee that the 
WS optical potential parameters listed in Table~\ref{tab:WS} result in as good 
fit to the data as provided by the GiBUU calculation.

One could also notice from Fig.~\ref{fig:sigabs} (left panels), 
that the BNL \cite{Abrams:1972ab} and Serpukhov \cite{Denisov:1973zv} data at 
$p_{\rm lab}=1.6\div20$ GeV/c are consistent with $\xi=1$, i.e. with the G-parity 
value of the real part of $\bar p$ optical potential.
A similar result based on the data \cite{Abrams:1972ab} has been obtained earlier in Ref. 
\cite{Bouyssy:1982cn}. The G-parity motivated $\bar p$ potential is, however, not 
supported by more recent low energy data of Ref. \cite{Nakamura:1984xw}. 
It would be, therefore, quite useful to perform the new measurements of $\bar p$ absorption
cross sections above 1 GeV/c, which is accessible at the future Facility for Antiproton and 
Ion Research (FAIR).

\begin{figure}
\includegraphics[scale = 0.50]{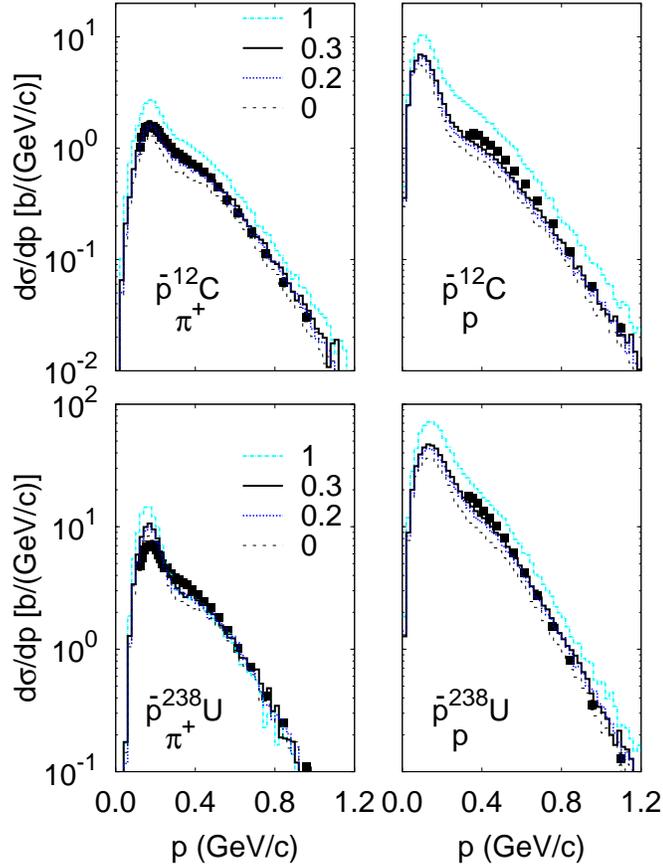}

\vspace*{1cm}

\caption{\label{fig:p_spectra}(Color online) The angle-integrated $\pi^+$ and proton 
laboratory momentum inclusive spectra from $\bar p$ interaction with $^{12}$C and 
$^{238}$U at 608 MeV/c. Calculated histograms are denoted by the value of the scaling 
factor $\xi$. Data are from \cite{Mcgaughey:1986kz}.} 
\end{figure}

As shown in Fig.~\ref{fig:p_spectra}, further constraints on the $\bar p$ optical potential 
can be obtained from inclusive momentum spectra of positive pions and protons produced in 
$\bar p$+$^{12}$C and $\bar p$+$^{238}$U interactions at 608 MeV/c. 
The two-slope structure of the pion spectra with the slope change at $p\simeq0.3$ GeV/c
is caused by pion-nucleon rescattering mediated by the $\Delta(1232)$ resonance. 
Higher momentum pions leave the nucleus practically without interactions with 
nucleons. Lower momentum pions are either absorbed via $\Delta(1232)$ resonances, 
$\Delta N \to N N$, or get decelerated in collisions with nucleons.
 
The calculated proton spectra also change their slopes at $p \simeq p_F$, where
$p_F=0.27$ GeV/c is the Fermi momentum of nucleons. High momentum protons are knocked-out
from a nucleus by energetic pions. The lower part of a proton momentum spectrum is populated
by the slow evaporated protons produced after the fast cascading pions and nucleons have 
already left the nucleus.

Varying parameters of the antiproton mean field influences the momentum spectra  
of annihilation products only moderately. The attraction of incoming $\bar p$ to 
a nucleus increases the annihilation probability. On the other hand, the invariant energy
of an annihilating $\bar p N$ pair is reduced by a stronger $\bar p$-attraction.
As a consequence, the multiplicities and kinetic energies of annihilation 
mesons get reduced. A partial cancellation of these two effects leads to a rather weak
sensitivity of the momentum spectra to the $\bar p$ mean field.
Overall, the calculation with $\xi=0.3$ is in the best agreement
with $\pi^+$ and proton momentum spectra at 608 MeV/c.

In conclusion, we have applied the hadron transport GiBUU model to describe 
$\bar p$-nucleus interactions at the beam momenta in the range $0.4\div280$ GeV/c.
The depth of the real part of the $\bar p$ optical potential, extracted by fitting the 
KEK data \cite{Nakamura:1984xw} at low beam momenta is $V_0\simeq150\pm30$ MeV which is 
about 40\% deeper than the value reported by Friedman et al \cite{Friedman:2005ad}.
However, depths corresponding to different geometries -- size
parameters -- may not be directly comparable with each other.

The annihilation spectra of positive pions and protons measured at LEAR \cite{Mcgaughey:1986kz} 
favour even deeper real part, $V_0\simeq220\pm70$ MeV.
Such attractive potentials may lead to the cold compression effect when $\bar p$ penetrates
deeply into the nuclear interior 
\cite{Buervenich:2002ns,Mishustin:2004xa,Larionov:2008wy,Mishustin:2008ch}.
This possibility deserves further studies in view of future FAIR experiments with antiproton beams. 

\begin{acknowledgments}
The support by the Frankfurt Center for Scientific Computing is gratefully 
aknowledged. We thank O. Buss, M. Kaskulov and L.M. Satarov 
for stimulating discussions. We are also indebted to T. Gaitanos 
for his permission to use the RMF implementation of the $\rho$-meson and Coulomb 
fields in the GiBUU model before publication.   
We are grateful to the referee for the productive exchange 
of ideas leading to the modified discussion.
This work was (financially) supported by the Helmholtz International
Center for FAIR within the framework of the LOEWE program (Landesoffensive 
zur Entwicklung Wissenschaftlich-\"Okonomischer Exzellenz) launched by the 
State of Hesse, by the DFG Grant 436 RUS 113/957/0-1 (Germany), and by the Grants 
NS-3004.2008.2 and RFBR-09-02-91331 (Russia). 
\end{acknowledgments}

\bibliography{pbarNuc_2}

\begin{figure}
\includegraphics[scale = 0.35]{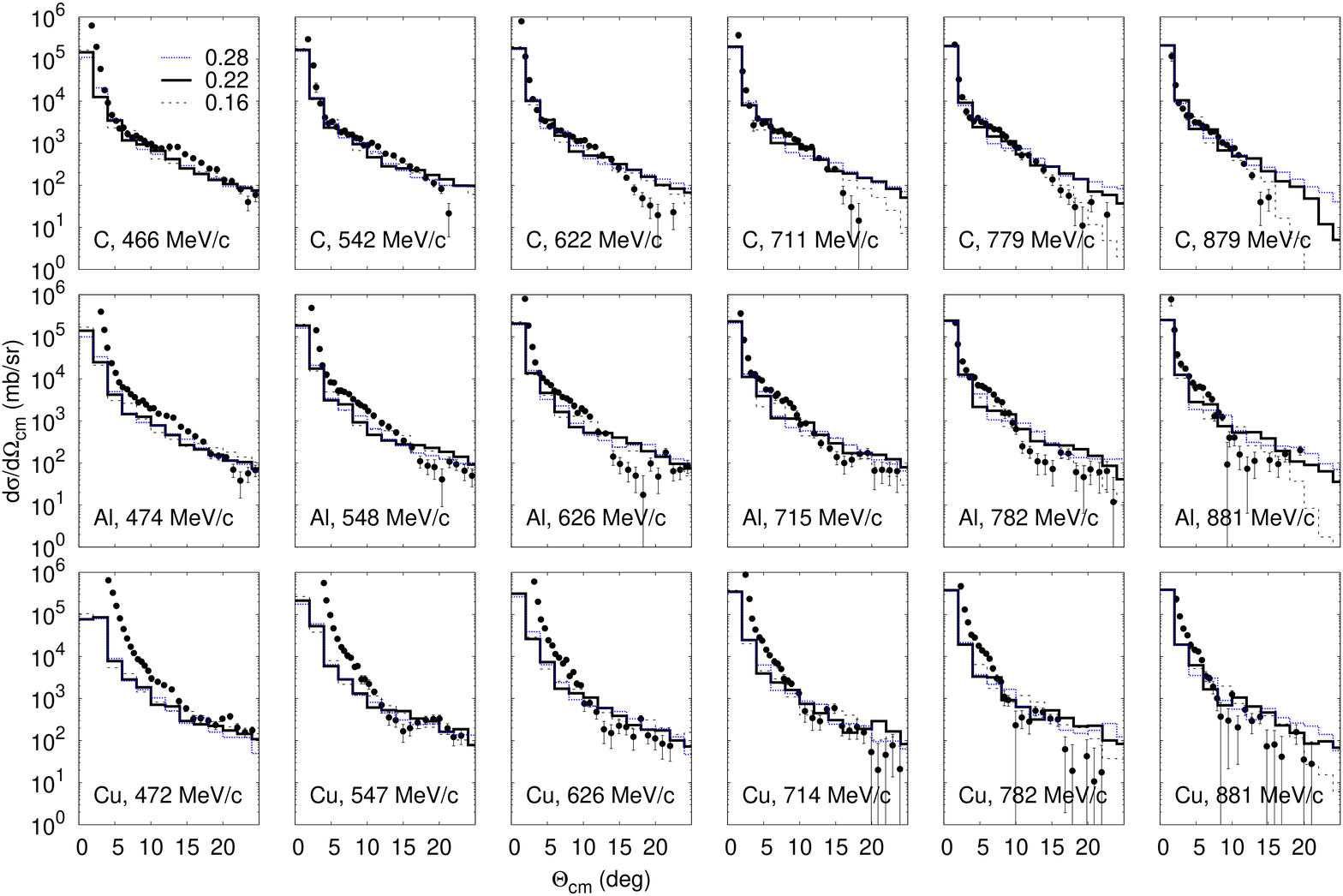}

\vspace*{1cm}

\caption{\label{fig:ELAST} (addition to the PRC version of the paper) 
$\bar p$ differential elastic scattering cross section. 
The computed histograms are marked with the value of a scaling factor $\xi$. 
Data are from Ref. \cite{Nakamura:1984xw}.
With some reservations for the semiclassical nature of our model, 
the agreement with the data is fairly good within the values of the scaling factor 
$\xi=0.22\pm0.06$, except for small-angle scattering at lower beam momenta, where we 
underpredict the experiment rather substantially. 
This deficiency is, most probably, due to neglecting the Coulomb corrections on the 
distant parts of the $\bar p$ trajectory (more than $\simeq(1.1A^{1/3}+5)$ fm from 
the nuclear centre). The detailed study of elastic scattering goes, however, 
beyond the scope of the present work.} 
\end{figure}

\begin{figure}
\includegraphics[scale = 0.6]{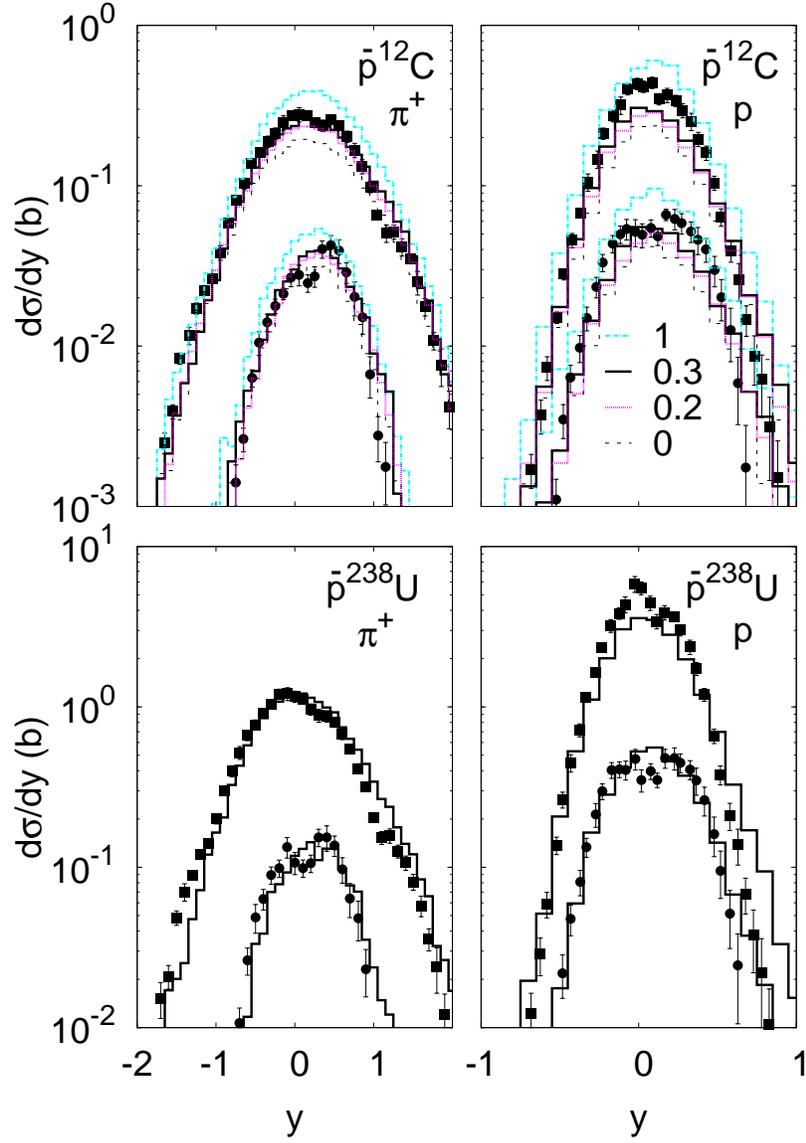}
\caption{\label{fig:y_spectra} (addition to the PRC version of the paper)  
$\pi^+$ and proton laboratory rapidity
spectra from antiproton interaction with $^{12}$C
and $^{238}$U at 608 MeV/c.
Calculated histograms are denoted by the value of the scaling factor $\xi$. 
For the $\pi^+$ (proton) spectra, the  upper lines and data points correspond to the transverse momenta 
$p_T \geq 120~(330)$ MeV/c, while lower lines and data points are computed for $p_T \geq 500~(600)$ MeV/c.
Experimental data are from \cite{Mcgaughey:1986kz}.}
\end{figure}

\end{document}